\documentclass{ifacconf}

\usepackage{graphicx}      
\usepackage{natbib}        
\usepackage{xcolor}
\usepackage{amsmath}
\usepackage{amssymb}
\usepackage[outdir=./]{epstopdf}
\begin{document}
\begin{frontmatter}

\title{Kalman-based interacting multiple-model wind speed estimator for wind turbines}


\author[First]{Wai Hou Lio} 
\author[First]{Fanzhong Meng} 

\address[First]{Department of Wind Energy, Technical University of Denmark, DK-4000 Roskilde, Denmark (e-mail: wali@dtu.dk, famen@dtu.dk).}


\begin{abstract}                

The use of state estimation technique offers a means of inferring the rotor-effective wind speed based upon solely standard measurements of the turbine. For the ease of design and computational concerns, such estimators are typically built based upon simplified turbine models that characterise the turbine with rigid blades. Large model mismatch, particularly in the power coefficient, could lead to degradation in estimation performance. Therefore, in order to effectively reduce the adverse impact of parameter uncertainties in the estimator model, this paper develops a wind sped estimator based on the concept of interacting multiple-model adaptive estimation. The proposed estimator is composed of a bank of extended Kalman filters and each filter model is developed based on different power coefficient mapping to match the operating turbine parameter. Subsequently, the algorithm combines the wind speed estimates provided by each filter based on their statistical properties. In addition, the proposed estimator not only can infer the rotor-effective wind speed, but also the uncertain system parameters, namely, the power coefficient. Simulation results demonstrate the proposed estimator achieved better improvement in estimating the rotor-effective wind speed and power coefficient compared to the standard Kalman filter approach.
\end{abstract}


\end{frontmatter}

\section{Introduction}

Typically, wind speed conditions in wind turbines are obtained based upon the wind anemometer located at the top of the nacelle. The quality of such a wind measurement suffers from the tower shadow, wind shear variations, and rotational sampling effect. Moreover, it is virtually impossible to describe the effect of the wind speed across the entire rotor plane by a single point of measurement taken at the anemometer. Thus, the concept of rotor-effective wind speed was emerged. This fictitious wind speed is either defined as the weighted sum of the wind speeds across the rotor where their weights are related to the power coefficient along the blade (e.g. \cite{Hooft2004}), or averaged wind speeds across the rotor plane (e.g. \cite{Ostergaard07}). The knowledge of rotor-effective wind speed is particularly valuable for modern large turbines and is important for many purposes such as advanced control strategies, for example, gain-scheduling controller (See \cite{LEITH1996}) and feed-forward control (See \cite{Meng2016}), health monitoring (e.g. \cite{Meng2018b}), power reserve estimation (e.g. \cite{Lio2019}) and down-regulation control (e.g. \cite{Lio2018a}). To obtain the rotor-effective wind speed, some studies suggested the use of LIDAR (LIght Detection and Ranging) systems could provide measurements of far upcoming wind speed but they are relatively expensive (See \cite{Schlipf}). In contrast, some articles demonstrated the use of state observers/estimators that infers the rotor-effective wind speed based on some existing standard measurement sensors, namely, generator speed and controller inputs.

In the literature, many state estimation methods have been applied to the rotor-effective wind speed estimation problem. For example, in \cite{Ma1995}, linear and extended Kalman filters were proposed to infer the rotor-effective wind speed using standard measurements of the turbine. In \cite{Ostergaard07}, a state observer with proportional-integral controller was employed to estimate the aerodynamic torque first and subsequently, based upon the torque estimate, the effective wind speed was inferred via inversion of the aerodynamic torque model. A study by \cite{Knudsen2011} included the turbine tower and induction dynamics in an extended Kalman filter design. Besides the standard observer, linear and exteneded Kalman filters, more state estimation techniques were studied such as unknown input observer (e.g. \cite{Odgaard2008}), disturbance estimator (e.g. \cite{Wright2004}), and immersion and invariance estimator (e.g. \cite{Ortega2013}). Detailed surveys and comparisons of rotor-effective wind speed estimation can be found in \cite{Soltani2013}.


Typically, the wind speed estimator is designed using a single state observer/Kalman filter for a given simplified turbine model. However, for superior performance, good knowledge of the model parameters is often required. In many studies, a simplified drive-train model was employed and its aerodynamic torque on the blade was characterised with the static power coefficient, that was obtained from simulating a blade with static forces. In reality, the power coefficient vary significantly dependent on the operating wind conditions and are difficult to be described simply by a simple static mapping (e.g. \cite{Jin2010}). In particular, for turbines operating in the above-rated wind conditions, the power coefficient changes significantly. Thus, a standard Kalman filter approach based on a simplified turbine model with a static power coefficient mapping tends to result in poor estimation performance in the above-rated wind region. 





Therefore, this paper develops a wind speed estimation algorithm that takes into account the model parameter uncertainties in the turbulent wind environment. In particular, the proposed estimator is designed based on the concept of interacting multiple-model (IMM) adaptive estimation (See \cite{Blom1988}). The proposed estimation strategy uses a parallel bank of extended Kalman filters to provide multiple estimates, where each filter is developed based on a different power coefficient mapping in various operations. Subsequently, the state estimate of the proposed estimator is computed based on a sum of each filter's estimate weighted by the likelihood of the filter model conditioned on the measurement. The adaptive nature of the proposed IMM estimator is important from an industry perspective, since not only the rotor-effective wind speed can be estimated based on solely standard turbine measurements, but also the IMM estimator can adapt itself to the true parameter of the operating turbines.


The reminder of this paper is as follows. Section~\ref{sec:2} presents the control-oriented modelling of wind turbines and motivation. The design of the proposed Kalman-based IMM estimator is shown in Section~\ref{sec:3}. In Section~\ref{sec:4}, the performance of the proposed estimator are demonstrated in simulation upon a high-fidelity and non-linear wind turbine model. It is followed by conclusions and future work in Section~\ref{sec:5}.

\section{Problem formulation}\label{sec:2}
\subsection{Modelling of wind turbines}\label{sec:2.1}

Typically, model-based state estimation requires a simplified model of the nonlinear system. The simplified model needs to capture the key dynamics of the turbine. In this study, a nonlinear turbine system model, that includes the dynamics of the rotor drive-train~\eqref{eq:rotor_dynamics} and wind speed~\eqref{eq:wind_dyn}, is employed. 

Firstly, the equation of motion of the drive-train is defined as follows:
\begin{subequations}\label{eq:rotor_dynamics}
\begin{align}
    J\dot{\omega} = \tau_\mathrm{a}(\lambda,\theta) -\tau_\mathrm{g}, 
\end{align}
where  $\omega\in\mathbb{R}$ is the rotor speed. The moment of inertia of the drive-train is denoted as $J \in \mathbb{R}$, the aerodynamic and generator torques are $\tau_\mathrm{a}: \mathbb{R}\times \mathbb{R} \rightarrow \mathbb{R}$ and $\tau_\mathrm{g}\in\mathbb{R}$, respectively, whilst $\theta\in\mathbb{R}$ is the pitch angle of the blades and the tip-speed ratio $\lambda\in\mathbb{R}$ is defined as follows: 
\begin{align}
    \lambda = \frac{\omega R}{ v},
\end{align}
where $R\in\mathbb{R}$ denotes the turbine blade length and $ v\in\mathbb{R} $ is the wind speed. The aerodynamic torque in~\eqref{eq:rotor_dynamics} is modelled as follows:
\begin{align}\label{eq:aeroTorq}
    \tau_\mathrm{a}(\lambda,\theta) = \frac{1}{2} \rho \pi r^2 C_\mathrm{p}(\lambda,\theta ) v^3 \omega^{-1},
\end{align}
where $r, \rho \in \mathbb{R}$ are the rotor radius and air density. The uncertainties in the power coefficient $C_\mathrm{p}(\lambda,\theta )$ is discussed in Section~\ref{sec:2.2}.
\end{subequations}

Secondly, the dynamics of the wind speed $v_k \in\mathbb{R}$ is assumed to be a random walk process and driven by a zero-mean white noise, defined as follows:  
\begin{align}\label{eq:wind_dyn}
    {v}_{k+1}  = {v}_k+n_k,~~n_k \sim \mathcal{N}\left(0,\sigma_n^2\right),
\end{align}
where the subscript $k\in\mathbb{Z}^{*}$ denotes the sample time and $n_k$ is the white Gaussian noise with zero mean and standard deviation $\sigma_n \in\mathbb{R}$. Notice that the model~\eqref{eq:wind_dyn} is sufficient to capture the slow dynamics of the wind (See \cite{Selvam2007}).

Finally. the discrete-time nonlinear turbine model can be constructed based on~\eqref{eq:rotor_dynamics} and~\eqref{eq:wind_dyn}:
\begin{subequations}\label{eq:nonlinear_sys}
\begin{align}
    x_{k+1}& = f (x_k,u_k)+w_{\mathrm{n},k},\\
    y_k & = h(x_k,u_k)+v_{\mathrm{n},k},
\end{align}
\end{subequations}
where $x_k= [\omega_k,{v}_k]^T\in\mathbb{R}^{n_x}$ denotes the system state vector, whilst $u_k\in\mathbb{R}^{n_u},y_k\in\mathbb{R}^{n_y}$ are the system input and output vectors. The state transition and output functions are denoted as $f:\mathbb{R}^{n_x}\times\mathbb{R}^{n_u} \rightarrow \mathbb{R}^{n_x},h:\mathbb{R}^{n_x}\times\mathbb{R}^{n_u} \rightarrow \mathbb{R}^{n_y}$. The process and measurement noises are linearly added to the system and denoted as $w_{\mathrm{n},k}\in\mathbb{R}^{n_x},v_{\mathrm{n},k}\in\mathbb{R}^{n_y}$, respectively.

\begin{figure}
    \centering
    \includegraphics[width=0.48\textwidth]{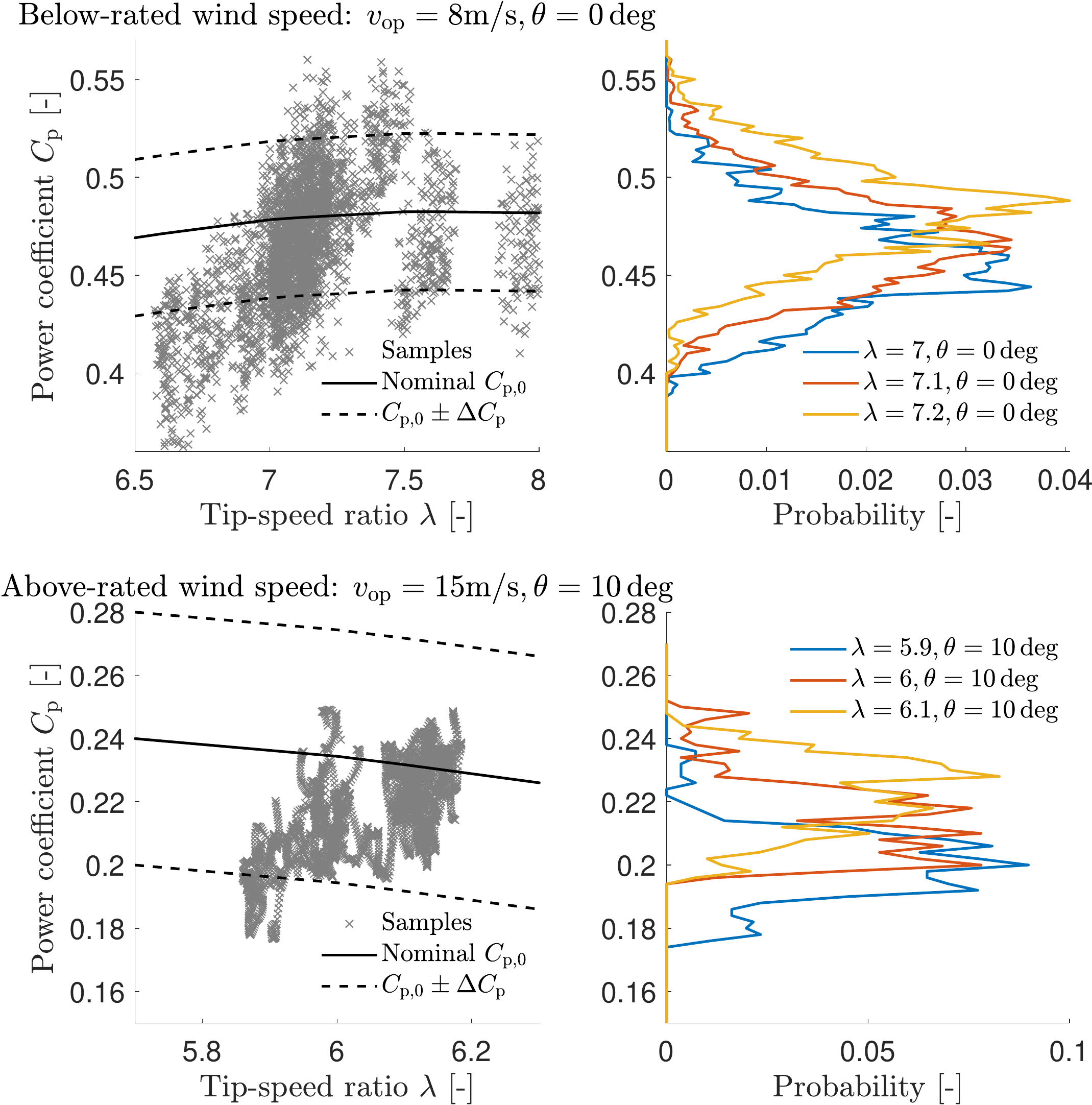}
    \caption{Left: Comparison between the static model prediction (solid line) and instantaneous $C_\mathrm{p}$ (crosses) generated in turbulent below- and above-rated wind cases. The dash lines represent the offset. Right: Probability distribution of the power coefficient $C_\mathrm{p}$ at specific tip-speed ratio and pitch angle in turbulent wind cases.  }
    \label{fig:VaryingCp}
\end{figure}


\subsection{Motivation and uncertainties in the power coefficient}\label{sec:2.2}

The simplified nonlinear wind turbine model~\eqref{eq:nonlinear_sys} presented in Section~\ref{sec:2.1} is a stochastic model assuming the process and measurement noise with zero means. The power coefficient $C_\mathrm{p}$ in aerodynamic torque~\eqref{eq:aeroTorq} is typically obtained based on a series of simulations calculating the equilibrium forces on the blade element (See \cite{Hansen}). The aerodynamic forces on the blades are computed based on different operating conditions, namely pitch angle, rotor speed and wind speed. Subsequently, these aerodynamic force are then integrated along the blade span yield the rotor-wise torques and finally, the power coefficient is obtained by multiplying these torques with the rotor speed and normalised with the available power in the wind. The aeroelastic code for generating the power and other coefficients is known as the blade element momentum (BEM) code. Typically, the power coefficient $C_\mathrm{p}$ in the aerodynamic torque~\eqref{eq:aeroTorq} is implemented in a look-up table or approximated analytical expression.

Nonetheless, the true power coefficient of an operating turbine is difficult to be characterised by a static mapping. The left columns of Figure~\ref{fig:VaryingCp} depicts typical power coefficient collected in dynamic turbulent wind field simulations in below- and above-rated wind conditions. These samples were then compared with the power coefficient computed using a static mapping (Nominal $C_\mathrm{p}$), obtained from the BEM code. The right columns of Figure~\ref{fig:VaryingCp} shows the distributions of the power coefficient samples at a specific operating conditions. Clearly, in the same operating conditions (pitch angle and tip-speed ratio), there exists a range of possible values for the power coefficient. The uncertainties in the model parameter $C_\mathrm{p}$ inevitably deteriorate the performance of the standard Kalman filter approach.



Therefore, the use of multiple-model approach is motivated in this work. A Kalman-based IMM estimator is proposed, where three Kalman filters are developed based on~\eqref{eq:nonlinear_sys} and each with a different $C_\mathrm{p}$ mapping, defined as follows.
\begin{align}\label{eq:cp_dev}
    C_\mathrm{p}(\lambda,\theta) = \begin{cases}
    C_\mathrm{p,0}(\lambda,\theta), & \mathrm{filter~1}\\
      C_\mathrm{p,0}(\lambda,\theta)+\Delta C_\mathrm{p}, & \mathrm{filter~2}\\
        C_\mathrm{p,0}(\lambda,\theta)-\Delta C_\mathrm{p}, & \mathrm{filter~3}.
    \end{cases}
\end{align}
where $C_\mathrm{p,0}$ denotes the nominal power coefficient look-up table that is calculated by the BEM code, whilst $\Delta C_\mathrm{p} \in\mathbb{R}$ denotes an offset that captures the uncertainty in the mapping, as shown in Figure~\ref{fig:VaryingCp}.


\section{Kalman-based IMM estimator } \label{sec:3}

This section presents the design of the proposed Kalman-based IMM estimator. The extended Kalman filter inside the IMM estimator is discussed in Section~\ref{sec:3.1} and followed by the IMM estimator design in Section~\ref{sec:3.2}.


\subsection{Extended Kalman filter}\label{sec:3.1}


A Kalman filter is a computationally efficient and recursive algorithm that provides the optimal state estimates $\hat{x}_k \in \mathbb{R}^{n_x}$ of a linear system by minimising the mean square state error, also known as the state error covariance matrix $P_k:=\mathbb{E}\left( (x_k - \hat{x}_k) (x_k - \hat{x}_k)^T \right)$. Since the turbine model in this work is a nonlinear model~\eqref{eq:nonlinear_sys}, an extended Kalman filter (EKF) is employed to estimate the states, namely, the rotor speed and wind speed. The EKF is similar to standard Kalman filter except that it computes the estimates based on the nonlinear equations and determines the state covariance matrix $P_k$ by linearising the system around the current state estimate. 

Typically, an EKF design consists of two phases: prediction update and measurement update. The superscripts $x_k^+, x_k^-$ are denoted as the \textit{a prior} and \textit{a posteriori} estimate, namely the estimate $x$ at sample time $k$ before and after the phase of measurement update, respectively. The discrete time EKF based on the model in~\eqref{eq:nonlinear_sys} is defined as follows:
\begin{subequations}\label{eq:ekf}

\begin{align}
    &\mathrm{Prediction~phase}:  \nonumber \\
    &\hat{x}^-_k = f(\hat{x}^+_{k-1},u_k), ~~  F_k := \frac{\partial f(\hat{x}^+_{k-1})}{\partial x},\\
    &P_k^- = F_k P^+_{k-1} F_k^T + Q_k, \\
    &\mathrm{Measurement~update~phase}:  \nonumber \\
    &\hat{y}_k = h(\hat{x}_k^-,u_k),~~\hat{x}_k^+ = \hat{x}^-_k + L_k (y_k - \hat{y}_k),   \\
    &P_k^+ = (I - L_k H_k) P^-_k,~~ H_k := \frac{\partial h(x_k^-) }{\partial x},
\end{align}
where $L_k\in\mathbb{R}^{n_x\times n_y}$ is the filter gain that minimises the mean square state error $P_k$ and it is computed as follows:
\begin{align}
    L_k = P^-_k H_k^T \left(H_k P^-_k H_k^T + R_k\right)^{-1},
\end{align}
\end{subequations}
where $Q_k\in\mathbb{R}^{n_x \times n_x},~R_k\in\mathbb{R}^{n_u \times n_u}$ are the covariance matrices of the process and measurement noises, respectively. They are typically computed as $Q_k = \mathbb{E}\left(w_{\mathrm{n},k}w_{\mathrm{n},k}^T\right)$ and $R_k = \mathbb{E}\left(v_{\mathrm{n},k}v_{\mathrm{n},k}^T\right)$.

\begin{figure}
    \centering
    \includegraphics[width=0.48\textwidth]{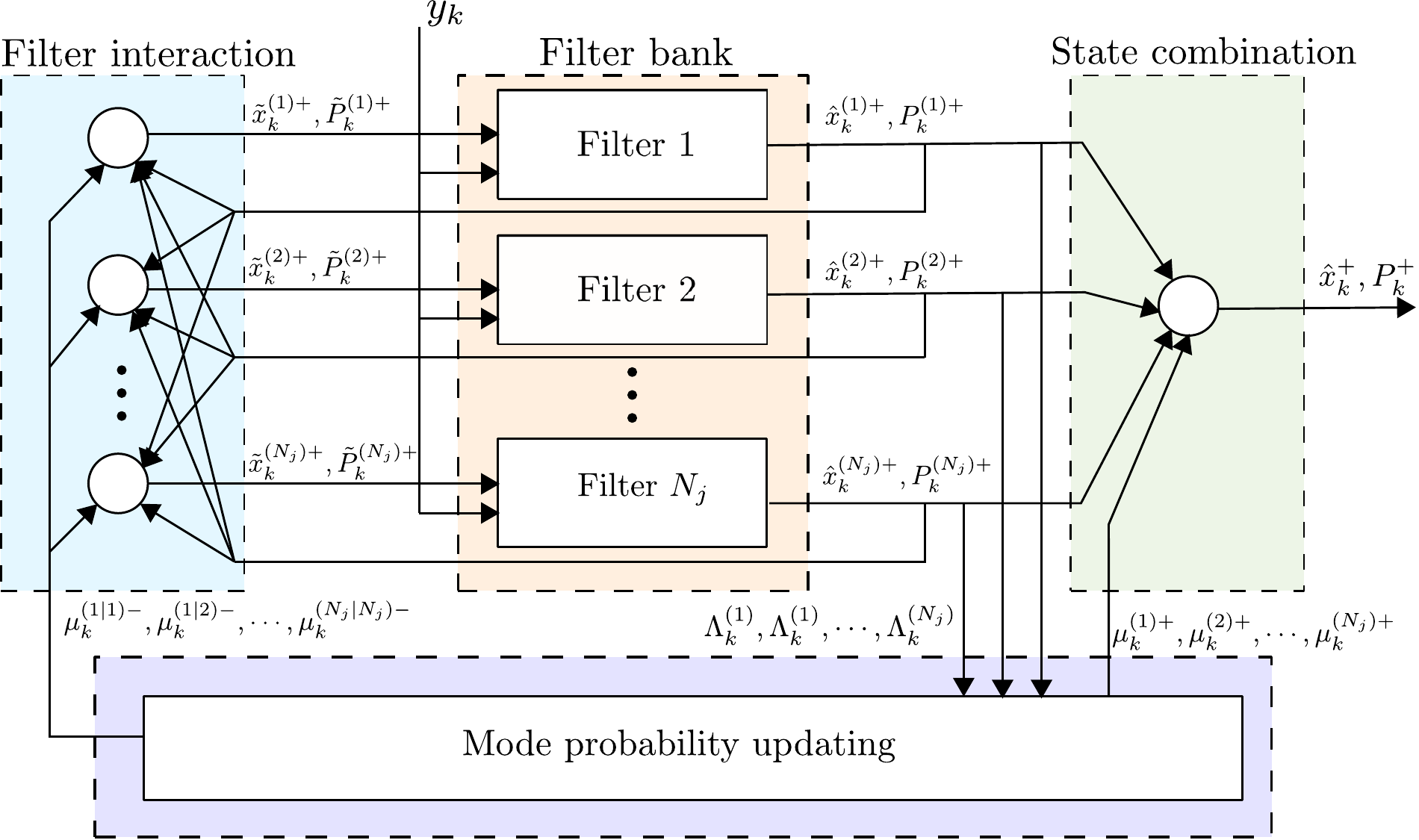}
    \caption{Architecture of the proposed interacting multiple-model (IMM) wind speed estimator.}
    \label{fig:imm}
\end{figure}

\subsection{Interacting multiple-model adaptive estimator}\label{sec:3.2}

The basic principle of an IMM estimator is to run a bank of filters corresponding to some unknown model parameters and subsequently, the estimates from each filter are fused based on the probability of each filter, which is typically calculated using their error residual and covariance. 

Figure~\ref{fig:imm} depicts the typical structure of an interacting multiple-model estimator. The algorithm mainly consists of four steps: filtering, mode probability updating, state combination and filter interaction. Notice the filtering step has been described in equation~\eqref{eq:ekf} in Section~\ref{sec:3.1}. Henceforth, the superscript $(j)$ indicates the index of the filter.

\subsubsection{Mode probability updating}
In IMM, a bank of EKFs is run in parallel to provide an estimate $x^{(j)+}_k$ and each filter has its own mode probability $\mu^{(j)-}_k$, where $j \in\mathbb{Z}$ is the filter number. These mode probabilities are updated from the filter likelihood, which is interpreted as how likely the filter provides a good state estimate from the measurement. Assuming the error residual is in Gaussian distribution, the likelihood is defined as follows:
\begin{align}
\Lambda^{(j)}_k = \frac{1}{\sqrt{2\pi \mathrm{det}(S_k^{(j)}) }} \mathrm{exp} (-\frac{1}{2} (z^{(j)}_k)^T (S^{(j)}_k)^{-1}  z^{(j)}_k  ),
\end{align}
where $z^{(j)}_k:= y_k-\hat{y}^{(j)}_k  \in \mathbb{R}^{n_y} $ is the residual whilst $S^{(j)}_k:=H^{(j)}_k P^{(j)-}_k (H^{(j)}_k)^T + R^{(j)}_k$ is the uncertainty (covariance) of the residual. Based on the Bayes' theorem, the mode probabilities are updated as follows:
\begin{align}
\mu^{(j)+}_k = \frac{\mu^{(j)-}_k  \Lambda^{(j)}_k}{\sum_j  \mu^{(j)-}_k  \Lambda^{(j)}_k}.
\end{align}

\subsubsection{State combination}
The output (state $\hat{x}^{+}_k$ and error covariance $P^{+}_k$) of the IMM estimator in Figure~\ref{fig:imm} are calculated by combining the weighted estimate $\hat{x}^{(j)+}_k$ and state error covariance $P^{(j)+}_k$ of each filter, defined as follows:
\begin{align}
\hat{x}^{+}_k &= \sum_j \mu^{(j)+}_k \hat{x}^{(j)+}_k, \\
P^{+}_k &= \sum_j \mu^{(j)+}_k  \left[  P^{(j)+}_k  + (  \hat{x}^{+}_k -  \hat{x}^{(j)+}_k ) (\hat{x}^{+}_k  - \hat{x}^{(j)+}_k )^T \right].
\end{align}

\begin{figure}
    \centering
    \includegraphics[width=0.48\textwidth]{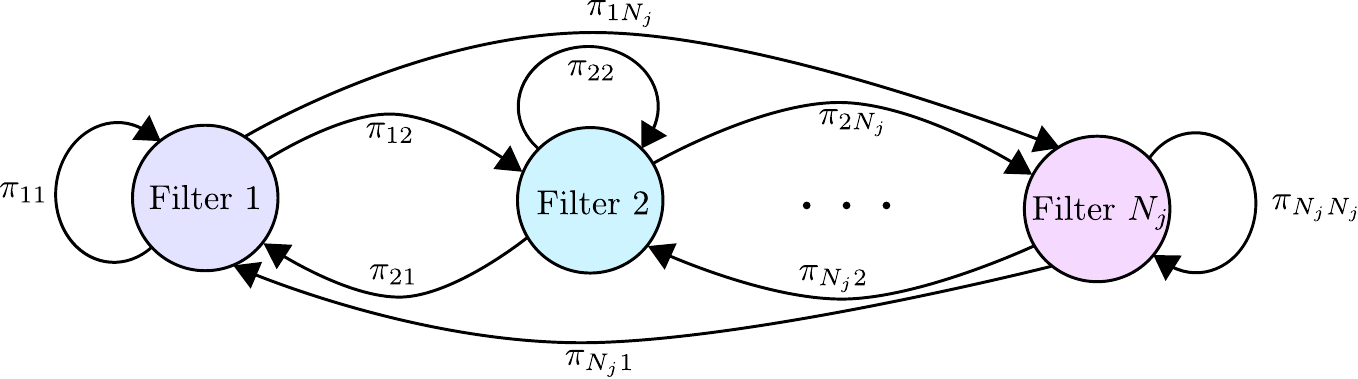}
    \caption{Markov switching model.}
    \label{fig:markov}
\end{figure}

\subsubsection{Filter interaction}


The filter interaction is that the filters with higher probabilities modify the estimates of the ones with lower probabilities. This step is particularly important for systems with time-varying uncertain parameters. The less likely filters are updated with better state and error covariance, thus, yielding a quicker response to changes in the model parameters. 

In IMM, each filter is considered as a mode and the switching process of the modes is modelled by a time-invariant Markov chain, depicted in Figure~\ref{fig:markov}. The mode transition probability $\pi_{ij}$ describes how likely the mode $i$ is changed to mode $j$ and notice that it is a design parameter. The priori mode probability for mode $j$ at next cycle $k+1$ is defined as follows:
\begin{align}
\mu^{(j)-}_{k+1}  = \sum_i \pi_{ij}  \mu^{(i)+}_{k}.
\end{align}
Subsequently, based on the Bayes' theorem, the mixing mode weight from mode $i$ to mode $j$ is calculated as follows:
\begin{align}
\mu^{(i|j)-}_k = \frac{\pi_{ij} \mu^{(i)+}_k }{\mu^{(j)-}_{k+1}} = \frac{\pi_{ij} \mu^{(i)+}_k} {  \sum_i \pi_{ij}  \mu^{(i)+}_{k}  }.
\end{align}
Finally, the mixed state $\tilde{x}^{(j)+}_k$ and mixed error covariance $\tilde{P}_k^{(j)+}$ are computed as follows:
\begin{align}
\tilde{x}^{(j)+}_k &= \sum_i \mu^{(i|j)-}_k \hat{x}_k^{(i)+},\\
\tilde{P}_k^{(j)+} &= \sum_i \mu^{(i|j)-}_k\\\nonumber
&\left[   P_k^{(i)+} + (  \tilde{x}^{(j)+}_k - \hat{x}^{(i)+}_k))   (  \tilde{x}^{(j)+}_k - \hat{x}^{(i)+}_k)^T         \right].
\end{align}

Now, the filtering step in equations~\eqref{eq:ekf} repeats with the mixed state $\tilde{x}^{(j)+}_k$ and error covariance $\tilde{P}_k^{(j)+}$ in the prediction update. The use of mixed state and error covariance ensure that the less likely filters are re-initialised with better estimates in the next step.

\section{Simulation results and discussions} \label{sec:4}

Simulation results are presented in this section.  The proposed Kalman-based IMM estimator was built with three EKF filters and each filter is based upon nonlinear turbine model~\eqref{eq:nonlinear_sys} with different power coefficient mappings as described in~\eqref{eq:cp_dev}. The proposed IMM estimator was compared against the standard single Kalman filter (KF), which was based upon solely the nominal power coefficient $C_{p,0}$ in~\eqref{eq:cp_dev}. The performance was examined based on estimation of the rotor-effective wind speed and power coefficient. The true rotor-effective wind speed was computed based on a averaged sum of 9 point measurements across the rotor. In terms of the simulation environment, the reference turbine model was the DTU10MW (See \cite{Bak2013a}) and the aeroelastic simulation code was the HAWC2~(See \cite{Larsen2007b}).  Two turbulent wind cases were considered: a mean wind speed of $v_\mathrm{op} = $ 8 m/s (below-rated) and 15 m/s (above-rated) with turbulent intensity of 10\%. The initial mode probabilities were equally distributed $\mu^{(1)}_0 = \mu^{(2)}_0 = \mu^{(3)}_0 =1/3$ and the mode transition matrix is chosen as follows:
\begin{align}
\Pi = \begin{bmatrix} 
\pi_{11} &\pi_{12} & \pi_{13}  \\ \pi_{21} & \pi_{22} &  \pi_{23} \\  \pi_{31} &  \pi_{32} &  \pi_{33}
\end{bmatrix} = \begin{bmatrix}
0.99 &0.005 & 0.005\\
0.005 & 0.99 & 0.005 \\
0.005 & 0.005 & 0.99
\end{bmatrix}.
\end{align}
This choice of the mode transition matrix was based on empirical judgement from a series of simulations. 

\subsection{Estimation of the  rotor-effective wind speed}\label{sec:4.2}
\begin{figure}
    \centering
    \includegraphics[width=0.48\textwidth]{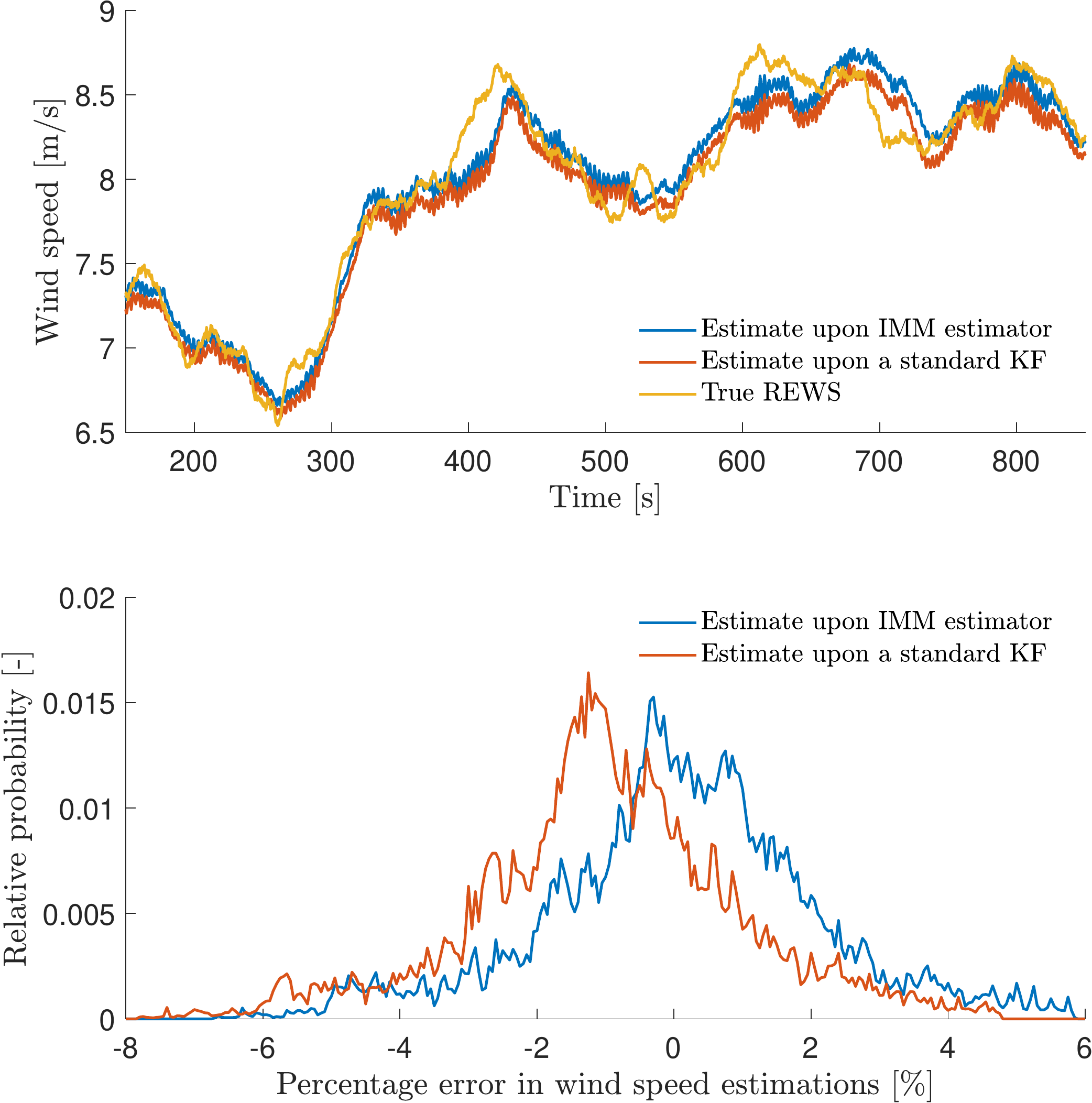}
    \caption{Top: Rotor-effective wind speed estimate of the proposed IMM estimator and standard KF in below-rated wind cases with a mean speed of $v_\mathrm{op}= 8 \mathrm{m/s}$. Bottom: Histograms of the estimation errors.}
    \label{fig:sim_v9}
\end{figure}

\begin{figure}
    \centering
    \includegraphics[width=0.48\textwidth]{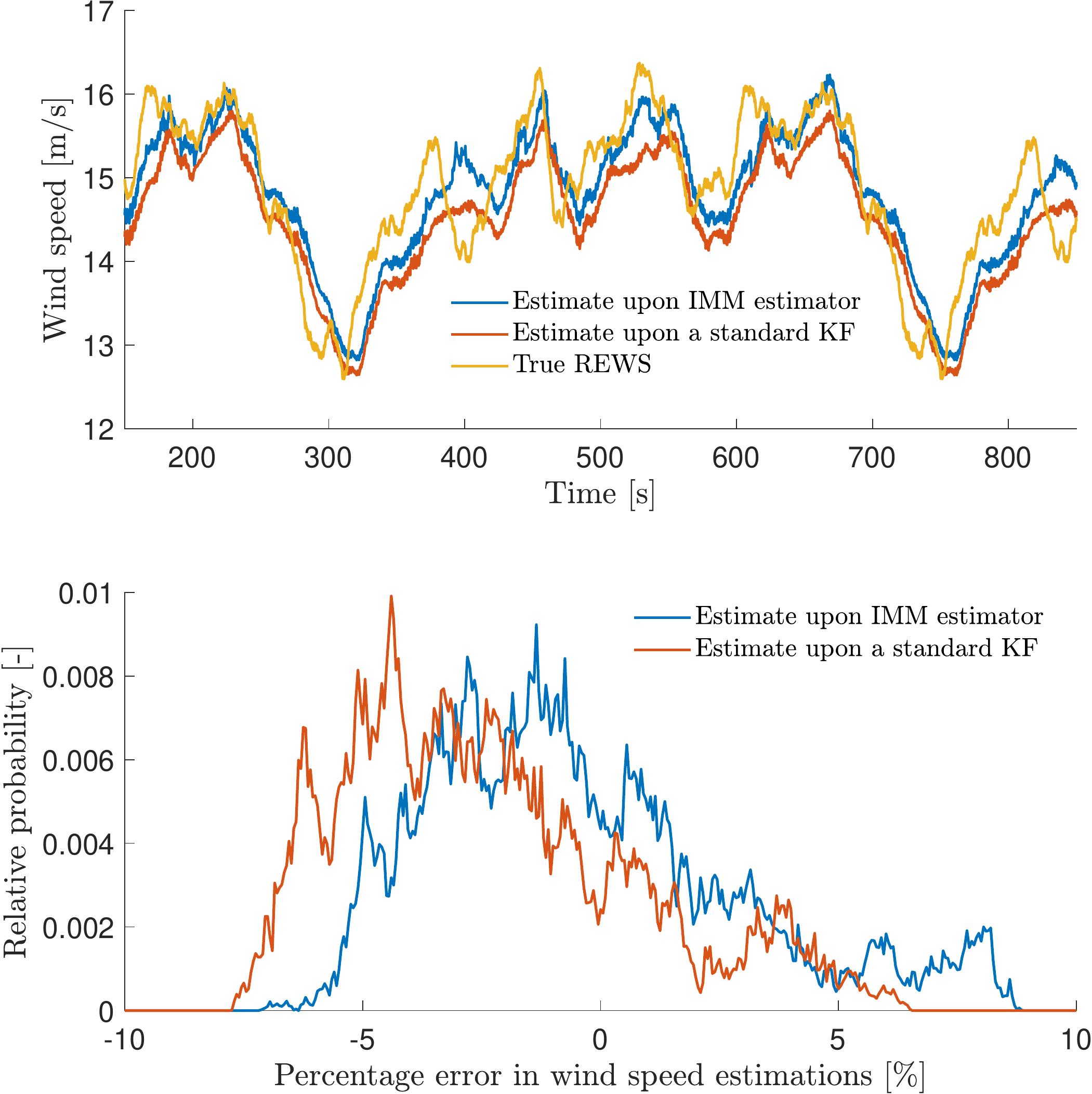}
    \caption{Top: Rotor-effective wind speed estimate of the proposed IMM estimator and standard KF in above-rated wind cases with a mean speed of $v_\mathrm{op}= 15 \mathrm{m/s}$. Bottom: Histograms of the estimation errors.}
    \label{fig:sim_v15}
\end{figure}

Figure~\ref{fig:sim_v9} and~\ref{fig:sim_v15} demonstrate the rotor-effective wind speed estimations under turbulent wind cases. Both estimators, the proposed IMM and standard KF, could achieve good estimation performance in both below-rated and above-rated wind conditions. However, the estimate from the proposed IMM was slightly better than the one from the standard KF, which is clearly shown in the corresponding histograms of the estimation errors. 



Table~\ref{tab:ws} summaries the means $\mu$ and standard deviations $\sigma$ of the errors. The mean errors of the proposed IMM estimator in both turbulent wind cases were significantly lower than the standard KF. The wind speed estimate is heavily dependent upon the power coefficient mapping. Particularly, in the above-rated wind region, the power coefficient tends to change frequently. The proposed Kalman-based IMM estimator considered the frequently changing model parameter, thus yielding a better estimation performance.





\begin{table}[]
\centering
\begin{tabular}{c|cc|cc}
                            & \multicolumn{2}{c|}{$v_\mathrm{op} = 8 \mathrm{m/s}$} & \multicolumn{2}{c}{$v_\mathrm{op} = 15\mathrm{m/s}$} \\ \cline{2-5} 
                            & $\mu$ {[}\%{]}        & $\sigma$ {[}\%{]}        & $\mu$ {[}\%{]}        & $\sigma$ {[}\%{]}       \\ \hline
IMM estimator & 0.08                 & -0.99                   & -0.42                 & -2.38                   \\
Standard KF   & 1.99                  & 1.95                     & 3.23                  & 3.02                   \\ \hline
\end{tabular}
\caption{Summary of estimation errors in the rotor-effective wind speed.}
\label{tab:ws}
\end{table}

\subsection{Estimation of the power coefficient}

\begin{figure}
    \centering
    \includegraphics[width=0.48\textwidth]{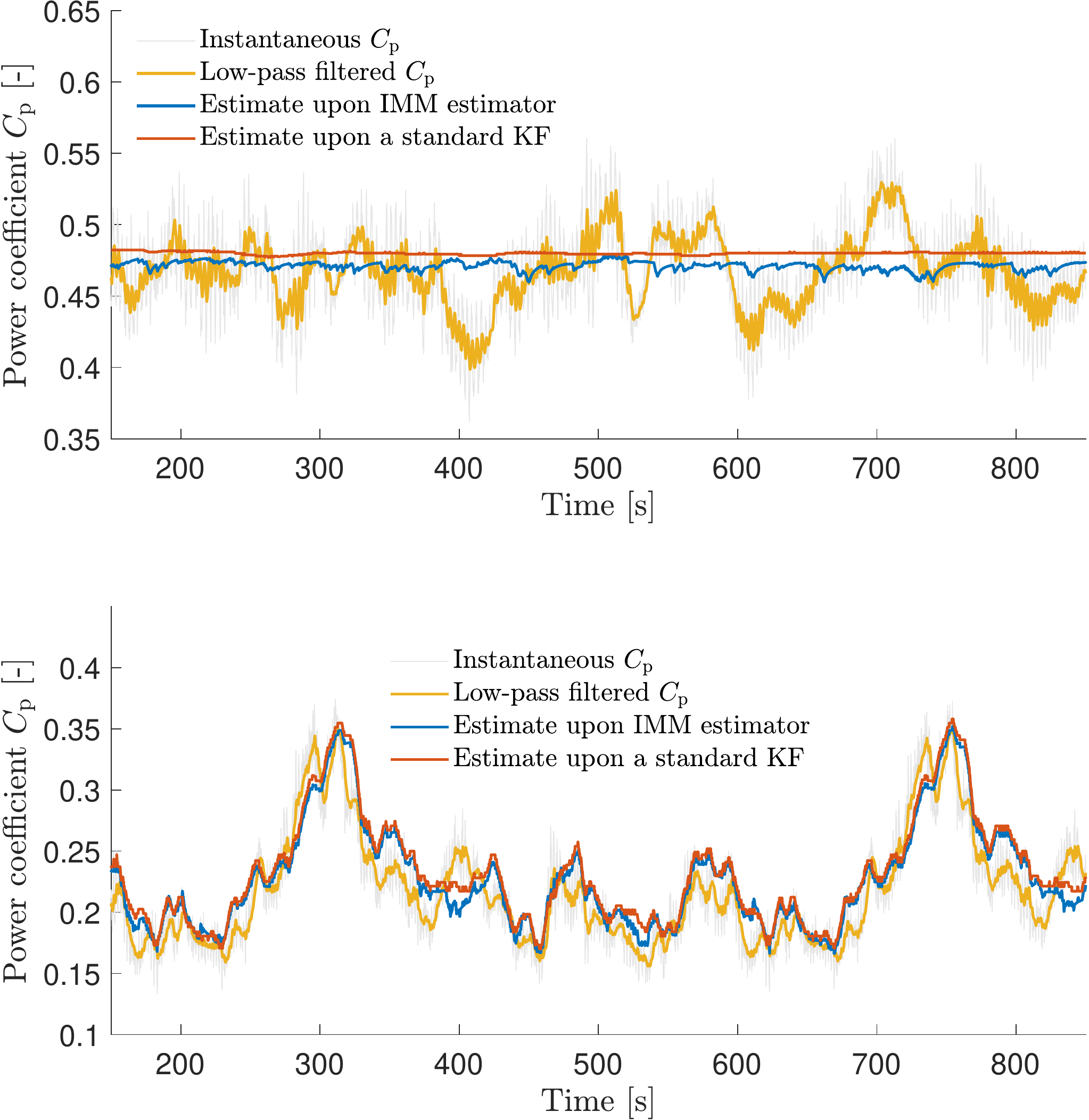}
    \caption{Estimates of the power coefficient by the proposed IMM and standard KF in turbulent wind cases with a mean speed of $v_\mathrm{op}=8\mathrm{m/s}$ (top) and $v_\mathrm{op}=15\mathrm{m/s}$ (bottom).}
    \label{fig:simCp}
\end{figure}

\begin{figure}
    \centering
    \includegraphics[width=0.48\textwidth]{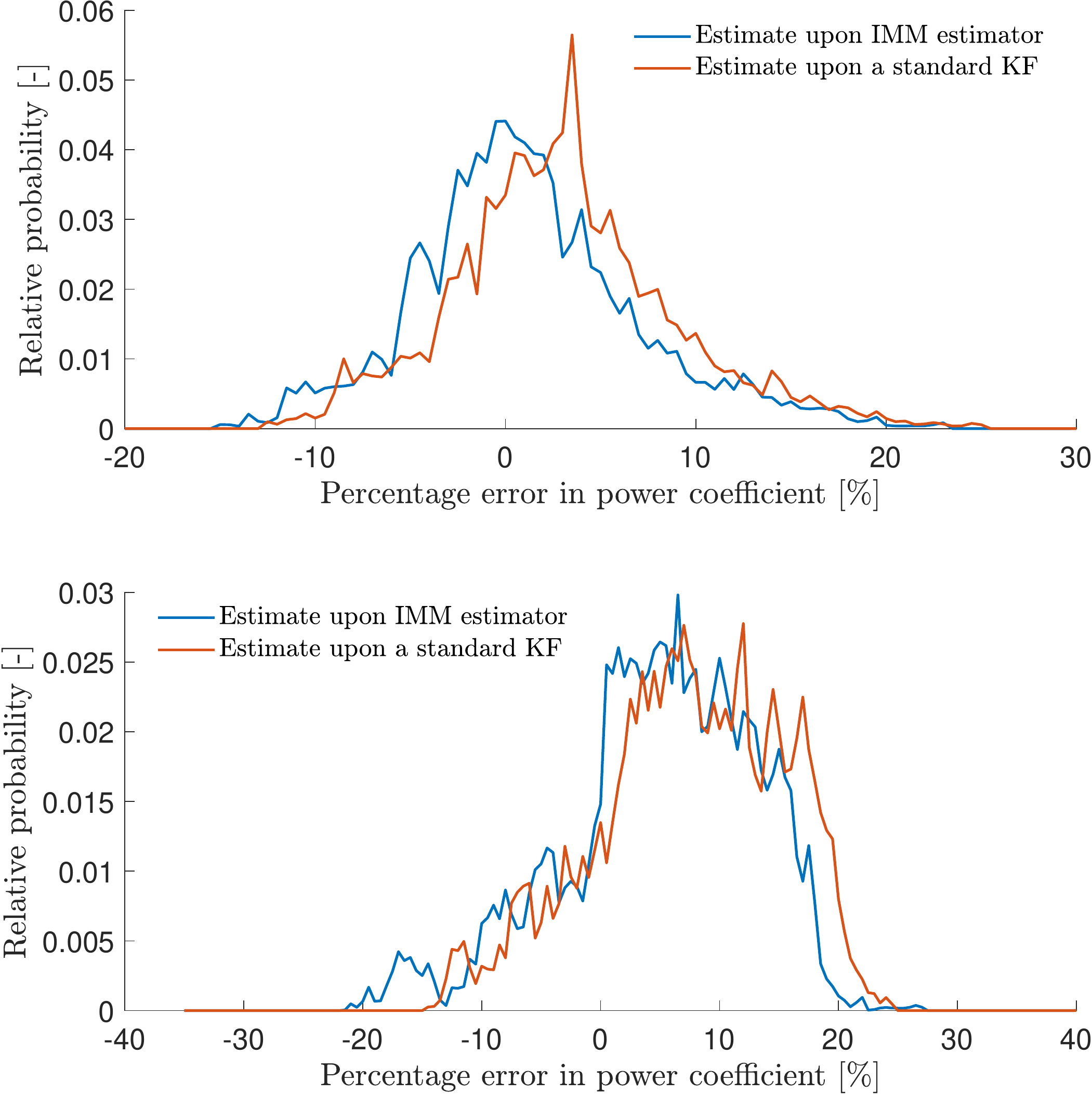}
    \caption{Histograms of the power coefficient estimation errors in turbulent wind cases with a mean speed of $v_\mathrm{op}=8\mathrm{m/s}$ (top) and $v_\mathrm{op}=15\mathrm{m/s}$ (bottom).}
    \label{fig:simCp_Hist}
\end{figure}

Besides the rotor-effective wind speed, the proposed IMM estimator and standard KF also could infer the power coefficient ($C_\mathrm{p}$). Figure~\ref{fig:simCp} shows the performance of the power coefficient estimation in both below-rated and above-rated turbulent wind cases.  Notice that the instantaneous power coefficient time-series contain many high frequency components that might not be useful for revealing the current turbine operating conditions, thus, a low-pass filtered $C_\mathrm{p}$ is also provided for comparisons with the cut-off frequency of 0.1 Hz. Generally, the $C_p$ estimation from the proposed IMM estimator performed better than the standard KF. This is confirmed by the histograms of the errors in Figure~\ref{fig:simCp_Hist}. Table~\ref{tab:cp} also summaries the means and standard deviations of the errors. Notice that the errors in Figure~\ref{fig:simCp_Hist} and Table~\ref{tab:cp} were calculated based on the difference between the estimate and low-pass filtered $C_\mathrm{p}$. In IMM estimator, the use of mode probability, that is calculated based on the state estimation error of each filter, reduced the bias in the power coefficient significantly.

\begin{table}[]
\centering
\begin{tabular}{c|cc|cc}
                            & \multicolumn{2}{c|}{$v_\mathrm{op} = 8 \mathrm{m/s}$} & \multicolumn{2}{c}{$v_\mathrm{op} = 15\mathrm{m/s}$} \\ \cline{2-5} 
                            & $\mu$ {[}\%{]}        & $\sigma$ {[}\%{]}        & $\mu$ {[}\%{]}        & $\sigma$ {[}\%{]}       \\ \hline
IMM estimator & 1.16                  & 3.04                    & 4.93                 & 7.30                  \\
Standard KF   & 5.90                  & 5.90                 & 8.07                  & 7.85                    \\ \hline
\end{tabular}
\caption{Summary of estimation errors in the power coefficient $C_\mathrm{p}$.}
\label{tab:cp}
\end{table}

\section{Conclusions and future works} \label{sec:5}

In this work, design of a Kalman-based interacting multiple-model (IMM) wind speed estimator was presented. The proposed IMM estimator was constructed based on three simplified turbine models with different power coefficient mappings. Simulation results were then presented that the proposed IMM estimator achieved better estimation performance compared to a standard Kalman filter. In particular, the biases in the estimation errors of the wind speed and power coefficient were significantly reduced. The use of multiple-model adaptive estimation technique is particularly relevant in turbine state observer design, since the blade aerodynamic parameters are changing frequently not only in operations, but also slowly over the turbine lifetime. Future work will look to use a time-varying mode transition matrix and additional set of power coefficient mappings.


\begin{ack}

 \begin{tabular}{ll}  
 \raggedleft
         \begin{tabular}{l}
          \includegraphics[width=0.08\textwidth]{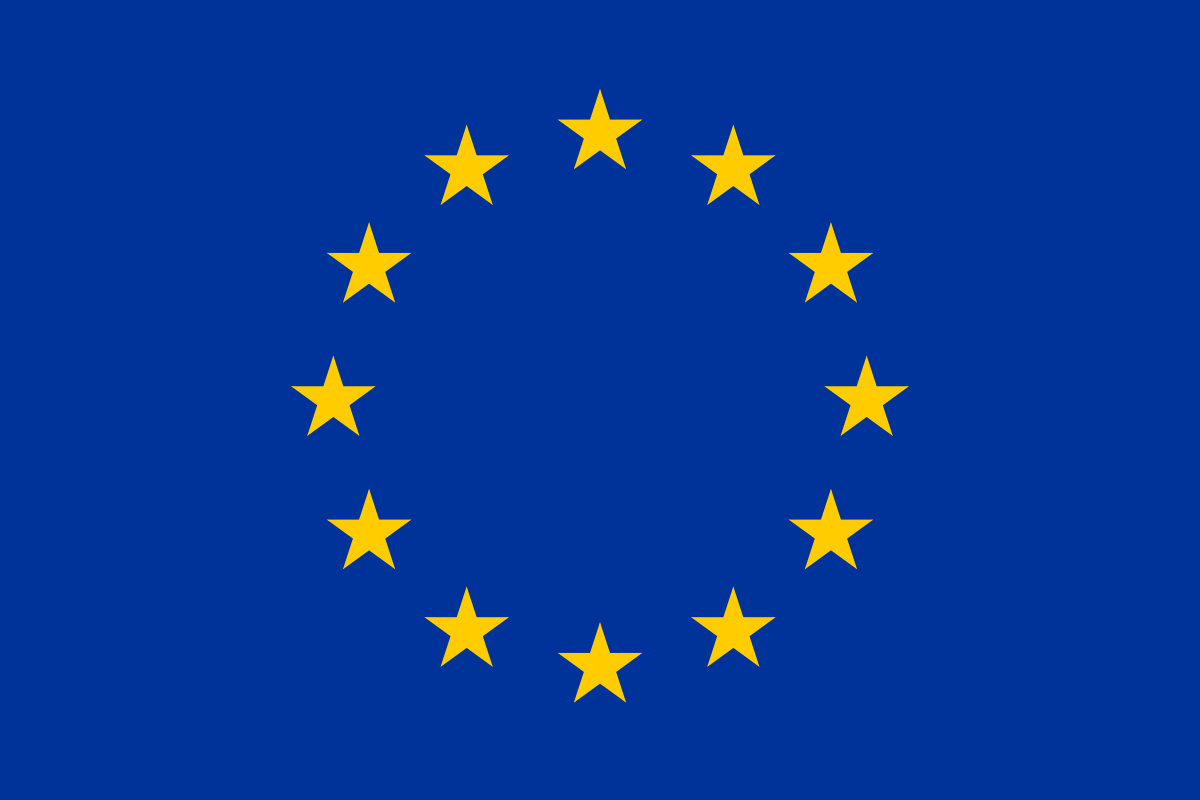}
           \end{tabular}
           & \begin{tabular}{l}
             \parbox{0.7\linewidth}{This study is supported by Danish Energy Agency (no. 64017-0045, PowerKey) and EU Horizon 2020 research programme (no. 727680, TotalControl).
    }
         \end{tabular}  \\
\end{tabular}
\end{ack}

\bibliography{ref.bib}             
\end{document}